\documentclass[aps,pre,twocolumn,showpacs,floatfix,superscriptaddress]{revtex4}
\usepackage{graphicx}
\usepackage{epsfig}

\begin{document}

\title{Complex Network Analysis of State Spaces for Random Boolean Networks}

\author{Amer Shreim}
\affiliation{Complexity Science Group, Department of Physics and Astronomy,
University of Calgary, Calgary, Alberta, Canada, T2N 1N4}

\author{Andrew Berdahl}
\affiliation{Complexity Science Group, Department of Physics and Astronomy,
University of Calgary, Calgary, Alberta, Canada, T2N 1N4}

\author{Vishal Sood}
\affiliation{Complexity Science Group, Department of Physics and Astronomy,
University of Calgary, Calgary, Alberta, Canada, T2N 1N4}
\affiliation{Institute for Biocomplexity and Informatics, University of
  Calgary, Calgary, Alberta, Canada, T2N 1N4}

\author{Peter Grassberger}
\affiliation{Complexity Science Group, Department of Physics and Astronomy,
University of Calgary, Calgary, Alberta, Canada, T2N 1N4}
\affiliation{Institute for Biocomplexity and Informatics, University of
  Calgary, Calgary, Alberta, Canada, T2N 1N4}

\author{Maya Paczuski}
\affiliation{Complexity Science Group, Department of Physics and Astronomy,
University of Calgary, Calgary, Alberta, Canada, T2N 1N4}

\date{\today}

\begin{abstract}
  We apply complex network analysis to the state spaces of random Boolean
  networks (RBNs). An RBN contains $N$ Boolean elements each with $K$ inputs.
  A directed state space network (SSN) is constructed by linking each
  dynamical state, represented as a node, to its temporal successor. We study
  the heterogeneity of an SSN at both local and global scales, as well as
  sample-to-sample fluctuations within an ensemble of SSNs. We use in-degrees
  of nodes as a local topological measure, and the path
  diversity~\cite{Shreim2007CA} of an SSN as a global topological measure.
  RBNs with $2 \leq K \leq 5$ exhibit non-trivial fluctuations at both local
  and global scales, while $K=2$ exhibits the largest sample-to-sample,
  possibly non-self-averaging, fluctuations. We interpret the observed ``multi
  scale'' fluctuations in the SSNs as indicative of the criticality and
  complexity of $K=2$ RBNs.  ``Garden of Eden'' (GoE) states are nodes on an
  SSN that have in-degree zero.  While in-degrees of non-GoE nodes for $K>1$
  SSNs can assume any integer value between $0$ and $2^N$, for $K=1$ all the
  non-GoE nodes in an SSN have the same in-degree which is always a power of
  two.

\end{abstract}

\maketitle

\section{Introduction}

In this paper we apply complex network analysis to discrete, disordered,
deterministic dynamical systems. The set of all trajectories of such a system
can be described as a directed network. Each dynamical state is represented by
a node and linked to its unique temporal successor by a directed link, giving
the state space network (SSN).  Thus the out-degree of a node is one.  The
irreversibility of the dynamics implies that a node can potentially have many
in-coming links.  The number of in-coming links at a node, or its in-degree,
can vary from node to node, depending on the dynamical system considered.  A
wide dispersity of degrees characterizes many complex
networks~\cite{NewmanSIAMReview}. Therefore, complex network analysis may
offer a useful alternative to traditional analyses of dynamical systems, such
as the characterization of spatiotemporal patterns~\cite{NKS,
  Peter1986IntJPhys, Badii, Bialek,Crutchfield}. The first results of such an
analysis on one dimensional cellular automata (CA) showed that heterogeneity
of the SSNs at both the local and global scales distinguishes ``complex''
dynamics from simple dynamics~\cite{Shreim2007CA}.  Here we exploit complex
network theory to characterize {\it disordered} dynamical systems by examining
SSNs for ensembles of random Boolean networks.

Random Boolean networks (RBNs) were introduced by Kauffman~\cite{Kauffman1969}
as models of gene regulation and have been extensively studied over the years
by physicists as examples of strongly disordered
systems~\cite{drosselRBNreview, aldanaRBN2003,
  derridaStauffer1986,flyvbjerg1988opn,BastollaParisi}.  The dynamics of each
of the $N$ Boolean elements in an RBN is given by a Boolean function of $K$
randomly chosen input elements. Different realizations of the input elements
and the Boolean functions for an RBN lead to an ensemble of RBNs for a given
$(N,K)$.

In the thermodynamic ($N\rightarrow \infty$) limit RBNs exhibit a phase
transition between chaotic and frozen phases passing through a critical phase
for $K=2$~\cite{DerridaPomeau}. In the frozen phase, $K<2$, the Hamming
distance between two perturbations of the same state quickly die out.  On the
other hand, in the chaotic phase, $K>2$, perturbations grow exponentially in
time.  Substantial analytical work has focused on the number of attractors as
well as their lengths~\cite{Drossel2005PRE, Drossel2005PRL,samuelsson2003RBN},
which have been found exactly for $K=1$ ~\cite{flyvbjerg1988esk}. Krawitz and
Shmulevich~\cite{ShmulevichEntropyRBN} computed the entropy of basin sizes and
showed that for critical RBNs it scales with the system size.  Otherwise, it
asymptotes to a constant.

In unrelated developments, a variety of statistical methods have been
established to characterize the structure of complex networks. Measures
include probability distributions for the node degree, clustering (the
tendency of nodes that share a common neighbor to also be linked to each other
directly), motifs (overrepresented subgraphs in the network),
etc.~\cite{NewmanSIAMReview, AlbertBarabasiRevModPhys, AlonMotifs2002}.  Many
real world networks such as regulatory networks~\cite{AlbertCellScience}, the
world-wide web~\cite{BarabasiWWW}, or the correlation structure of
earthquakes~\cite{MayaNetworksEarthquakes,MayaNetworksEarthquakes2}) differ
markedly from a random graph -- where the degree distribution is Poisson and
clustering is absent. They often display ``fat-tailed'' or even scale-free
degree distributions.

In this paper we show that strong deviations from random graph behavior also
occur in the SSNs of RBNs.  In particular, for sufficiently small $K$, the
probability distribution for the number of incoming links to nodes in the SSN,
$P(k)$, has a fat-tail. In contrast, a randomly rewired null model for the SSN
({\it i.e.}, a random map) has Poissonian in-degree distributions. Also for
small $K$, the maximum in-degree of any node in the SSN, $k_{max}$, displays
scaling behavior with respect to size of the SSN, ${\cal N}=2^N$.  While
$k_{max}$ is a measure of local heterogeneity, we use the path diversity,
${\cal D}$, to characterize global heterogeneity~\cite{Shreim2007CA}. We show
that ${\cal D}$ grows linearly with $N$ for $K=1$, while it grows faster than
linearly with $N$ for $2 \leq K \leq 5$. For $K\geq 6$, ${\cal D}$ scales with
SSN size ${\cal N}$.  In addition, for $K=2$ (where RBNs are critical) the
sample to sample fluctuations of ${\cal D}$ are the largest and might be
non-self-averaging~\cite{chamati2002fss, wiseman1998fss}.  We speculate that
SSN fluctuations at these three different scales (local, global, and sample to
sample) for $K=2$ RBNs are associated with criticality in the thermodynamic
limit ${\cal N} \rightarrow \infty$.

\subsection{Summary}

In Section~\ref{def} we discuss the procedure used to construct RBNs, and their
SSNs. The in-degree and the path diversity are also defined.  In
Section~\ref{indeg}, we present results for the behavior of the nodes'
in-degrees in an ensemble of SSN.  In Section~\ref{K1}, we discuss the SSNs
for $K=1$ RBNs, which have unique features not shared by other $K$s. In
Section~\ref{diversity}, we examine the behavior of the path diversity.
Discussion of our results and concluding remarks are found in
Section~\ref{discussion}.

\section{Definitions}
\label{def}

An RBN consists of $N$ Boolean variables $\sigma_i \in {0,1}$ with $ i=1 \cdots
,N$.  The dynamics of each element is determined by a Boolean function of $K$
randomly chosen input elements,
\begin{equation} 
  \sigma_i(t+1)   = f_i(\sigma_{i_1}(t),\sigma_{i_2}(t),\ldots,\sigma_{i_K}(t))  \quad ,
  \label{function}
\end{equation}
where $\sigma_{i_j}(t)$ is the value of the $j^{th}$ input to $\sigma_i$ at
time $t$. The function $f_i$ is randomly chosen to be $1$ with probability $p$
and $0$ with probability $1-p$ for each set of values of its arguments.  We
consider only the case of unbiased RBNs with $p=1/2$, except when $K=1$, where
we study two different cases. For $K=1$ there are four possible boolean
functions for each element. Instead of choosing them uniformly, using only the
copy ($f(\sigma) = \sigma$) or the invert function ($f(\sigma) = {\rm
  NOT}(\sigma)$) for each $\sigma_i$ (with equal probability), leads to a
\emph{critical} $K=1$ RBN~\cite{drosselRBNreview}.

We use synchronous update for the dynamics, {\it i.e.} all the Boolean
elements in the network are updated in parallel at each time step. The
networks are set up by choosing $K$ different random inputs for each
$\sigma_i$.  While allowing self-connections, we do not allow multiple
connections. We do not impose any connectivity constraint on the RBNs.

\begin{figure}[htbp]
\begin{center}
\includegraphics*[width=8cm]{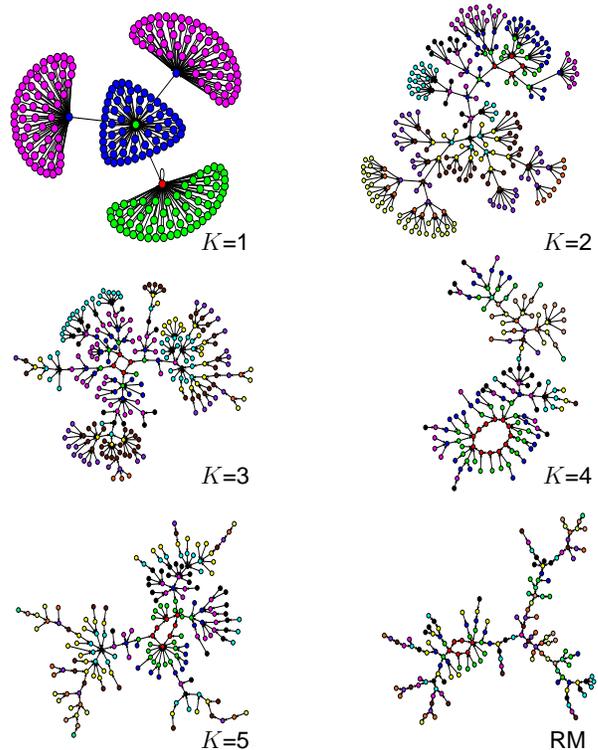}
\caption{\label{networks}(Color online).  One connected cluster of an SSN for
  RBNs with $N=9$ and different values of $K$ displayed using the program
  'Pajek'~\cite{Pajek}.  Nodes on the attractors are drawn in red; the other
  colors indicate distance from the attractor. For instance, for $K=1$, green
  nodes are distance one from the attractor, blue nodes are distance two and
  magenta nodes are distance three. Note that for $K=1$ all nodes are either
  Garden of Eden states or hubs with all the hubs having the the same
  in-degree.  RM stands for a random map.}
\end{center}
\end{figure}

An RBN with $N$ elements has ${\cal N} =2^N$ different dynamical states.
These states are nodes of a directed network.  A link from ${\bf A}$ to ${\bf
  B}$ indicates that ${\bf A}$ evolves to ${\bf B}$ in one time step, making
${\bf B}$ the image of ${\bf A}$, or ${\bf A}$ a pre-image of ${\bf B}$.  This
directed network forms the state space network (SSN).The SSN typically
consists of disconnected clusters, or basins of attraction.  Each basin
contains transient states, which are visited no more than once on any
dynamical trajectory, and attractor states that may be visited infinitely
often.  Garden of Eden states (GoE) are transient states that cannot be
reached from any other state, {\it i.e} they have no pre-images. Examples of
some state space clusters are shown in Fig.~\ref{networks} for $N=9$ and
different values of $K$.

Random maps are the limit of RBNs when $N,K \rightarrow
\infty$~\cite{DerridaFlyvbjerg}.  By construction, the state space of a random
map forms an Erd{\"o}s-Reyni random graph with a Poisson in-degree
distribution, $P(k)=e^{-1}/k!$ with mean $\langle k \rangle = 1$. We
constructed random map SSNs by picking the image of each node uniformly
randomly from the ${\cal N} =2^N$ possible nodes.

Identifying all attractors states and lumping them into a single node turns
the SSN into a rooted tree.  To study the global heterogeneity of this tree,
we use path diversity $\cal D$. It quantifies topological variations in the
paths connecting the GoE states to the root. It is similar to other global
measures of tree like structures such as tree diversity~\cite{Huberman} and
topological depth~\cite{Badii}. ${\cal D}$ measures the number of
non-equivalent choices encountered when following each reversed path from the
root to the GoE states.

Specifically, $\cal D$ is computed as follows: each GoE state is assigned a
diversity equal to one; a state with a single pre-image has the same diversity
as its unique pre-image; and the diversity of a node with more than one
incoming link is the sum of all \emph{distinct} diversities of its pre-images
plus one. For instance, if a node has in-degree 6, and three of its pre-images
have diversity 4, two have diversity 5, and the last has diversity 8, then
that node's diversity is $4+5+8+1=18$. Finally, the path diversity ${\cal D}$
of the entire SSN is the diversity of the root.

Table~\ref{TableOfNotation} provides a summary of the notation used. Note that
we refer to nodes of the RBNs as ``elements'' and otherwise reserve the phrase
``node'' only for SSNs.

\begin{table}[!h]
\begin{center}
\begin{tabular}{|l||l|}
\hline
RBN              & random Boolean network       \\
\hline
$N$              & number of elements of an RBN  \\
\hline
$K$              & number of inputs to each element of an RBN \\
\hline
SSN              & state space network of an RBN \\
\hline
$\cal N$         & number of nodes in an SSN (${\cal N}=2^N$) \\
\hline
$k$              & in-degree of a node in an SSN \\
\hline
$k_{max}$         & largest in-degree of a node in an SSN \\
\hline
$\cal D$         & path diversity \\
\hline
\end{tabular}
\caption{\label{TableOfNotation} Notations used in this paper.}
\end{center}
\end{table}

\section{Results}

\subsection{In-Degree}
\label{indeg}

An elementary description of local heterogeneity of a network is its degree
distribution.  We computed in-degree distributions, $P(k)$, of the SSNs. These
SSNs were obtained for RBNs with $1 \leq K \leq 10$ and $10 \leq N \leq 24$,
as well as random maps of system sizes $10 \leq N \leq 24$. For $K=1$ we
distinguish between RBNs constructed using all four boolean functions and the
critical $K=1$ RBN discussed earlier.  In Section ~\ref{K1} we discuss the
in-degrees for $K=1$ RBNs, where we obtain exact analytic results. We present
in Section ~\ref{K2} numerical results for $K>2$ RBNs.

\subsubsection{K=1 Networks}
\label{K1}

We start with the case of $K=1$ critical RBNs, where all functions
$f_i(\sigma_{i_1})$ are either copy or invert. If none of the boolean elements
is a leaf on the RBN, then the value of every element $\sigma_i$ at the next
time step is determined by the current value of of some element $\sigma_j$. If
the state $\bf{S}$ differs from $\bf{S}'$ in the value of $\sigma_i$, their
pre-images will differ in the value of $\sigma_j$.  Thus on a $K=1$ critical RBN
without leaves, the mappings are one-to-one, there are no GoE states and each
hub has in-degree one.

When the RBN contains $L$ leaves in addition to $Q$ non-leaf elements, the
state of the $L+Q$ elements can be written as a concatenation,
\begin{equation}
\bf{S} = \bf{L} \oplus \bf{Q} \label{concatenation}.
\end{equation}
If $\sigma_i$ is among the $L$ leaves, it cannot effect the next step value of
any element. The next value of $\sigma_i$ itself is determined by the current
value of a non-leaf element $\sigma_j$.  The image of $\bf{S}$ under the RBN
rules will then be of the form,
\begin{equation}
{\cal I}(\bf{S})={\cal F}(\bf{Q}) \oplus {\cal G}(\bf{Q}),
\end{equation}
where ${\cal F}$ gives the next-step state of the leaves, and is a function
only of the non-leaves. This means that at least $2^L$ states that differ only
in the values of the $L$ leaf elements will be mapped to the same image,
{\it i.e.} the in-degree of each non-GoE state (``hubs''), is $k_h \geq 2^L$.
To see that $k_h=2^L$, we note that 
\begin{eqnarray}
\label{notsame}
{\cal I}({\bf S}')\neq{\cal I}({\bf S})  {~\rm if}  {~\bf Q}'\neq {\bf Q}.
\end{eqnarray}
Consider two states ${\bf Q}$ and ${\bf Q}'$ of the non-leaf elements that
differ in the state of the element $\sigma_a$.  If $\sigma_a$ is an input to
at least one non-leaf element, then ${\cal G}(\bf{Q}')\neq{\cal G}(\bf{Q})$.
If $\sigma_a$ is an input to at least one leaf element, then ${\cal
  F}(\bf{Q}')\neq{\cal F}(\bf{Q})$.  Since $\sigma_a$ is not a leaf, at least
one of these two cases has to be true, proving Eq.~(\ref{notsame}).  Thus the
cardinality of the set $\{{\cal I}({\bf S})\}$ is at least equal to the
cardinality of the set $\{{\bf Q}\}$, which is $2^Q$. If for each hub
$k_h>2^L$, the sum of hub in-degrees $\geq 2^Q 2^L = 2^{Q+L} = 2^N$. Since
there are only $2^N$ nodes in the SSN, this inequality has to be saturated,
giving us that $k_h = 2^L$.

The boolean function used for $K=1$ critical RBNs are either copy or invert.
For the general $K=1$ case, there are four possible functions, two of which
are copy and invert. The other two are constant functions, which map all
aruguements to one value (either zero or one). If an element $\sigma_i$ is
input only to elements with constant functions, it behaves effectively like a
leaf. We will refer to such elements along with the leafs, as effective
leaves.  Using this observation we can extend the above argument for $K=1$
critical RBNs to the general case of $K=1$ RBNs. Thus all hubs in the SSN of a
$K=1$ RBN have the same in-degree of the form $k_h=2^l$, where $l$ is the
number of effective leaves.  However, $k_h$ will vary over the different
realizations of the RBN.

Construction of an instance of a $K=1$ RBN can be considered as $N$ rolls of
an $N$-faced die.  Each face represents an element. The outcome $j$ of the
$\tau$-th roll is the input to the $\tau$-th element of the RBN. The function
$f_{\tau}$ in Eq.~(\ref{function}) that determines the updates of the
$\tau$-th element is chosen randomly over the function's arguments, $\sigma_j$
in the present case.  Since for $K=1$ there is only one input, $f_\tau$ will
be a constant function if it maps both values of the input (0,1) to the same
value, either $1$ (with probability $p^2$) or $0$ (with probability
$(1-p)^2$).  Therefore, the probability that the chosen function is not a
constant function is $q=1-p^2 - (1-p)^2$. We flip a coin, after each roll, to
determine if the function $f_{\tau}$ is a constant function or not.  If it is
not a constant function, we mark the displayed face of the die, and keep track
of the number of marked faces until $\tau=N$.  The marked faces exclude the
candidates for effective leaves. After $N$ die rolls and coin flips, the
number of marked faces will be $m = N-l$, where $l$ is the the number of
effective leaves in the constructed RBN. The distribution of $m$ evolves from
the $\tau$-th to the $\tau+1$-th roll according to,
\begin{equation}
M_{\tau+1,m}= q\left(1-\frac{m-1}{N}\right)M_{\tau,m-1}+\left(1-q(1-\frac{m}{N})\right)M_{\tau,m},
\label{dist-marked}
\end{equation}
under the boundary conditions that
\begin{eqnarray*}
	M_{\tau,m}=0, {~\rm if} ~\ m<0 {~\rm or} ~\  m>\tau, 
\end{eqnarray*}
and the initial condition $M_{0,0}=1$.

 Thus $q=1$ in Eq.~(\ref{dist-marked}) generates the
distribution of leaves for $K=1$ critical RBNs. In this case, the solution to
Eq.~(\ref{dist-marked}) is
\begin{equation}
\label{q1hubexact}
	M_{N,m} = \frac{1}{N^N}{N\choose m}C_{N,m}m!,
\end{equation}
where $C_{N,m}$ are Stirling numbers of the second kind, and enumerate
partitions of an $N$-set into $m$ non-empty subsets \cite{abramowitz1965hmf}.
For $q\neq1$ analytical expressions are hard to obtain. However, solving
Eq.~(\ref{dist-marked}) numerically is straightforward.

\begin{figure}[htbp]
\begin{center}
	\includegraphics*[width=8cm]{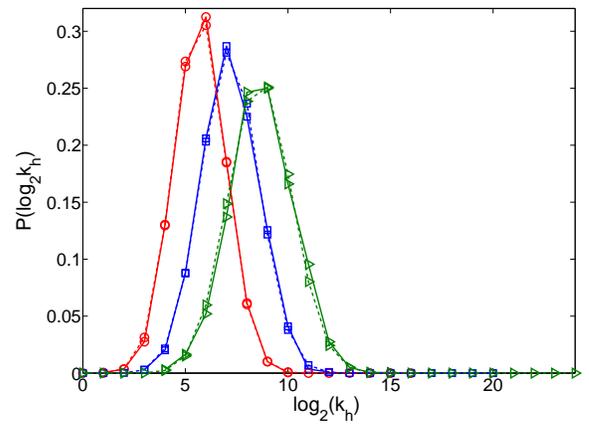}
        \caption{\label{PDFlogkmaxK1crit}(Color online). The PDF $P(\log_2 k)$
          of the log of the hub in-degree, $\log_2 k_h$ for $K=1$ critical RBNs.
          The three dashed lines with empty symbols were obtained using
          Eq.~(\ref{dist-marked}) with $q=1$.}
\end{center}
\end{figure}

\begin{figure}[htbp]
\begin{center}
	\includegraphics*[width=8cm]{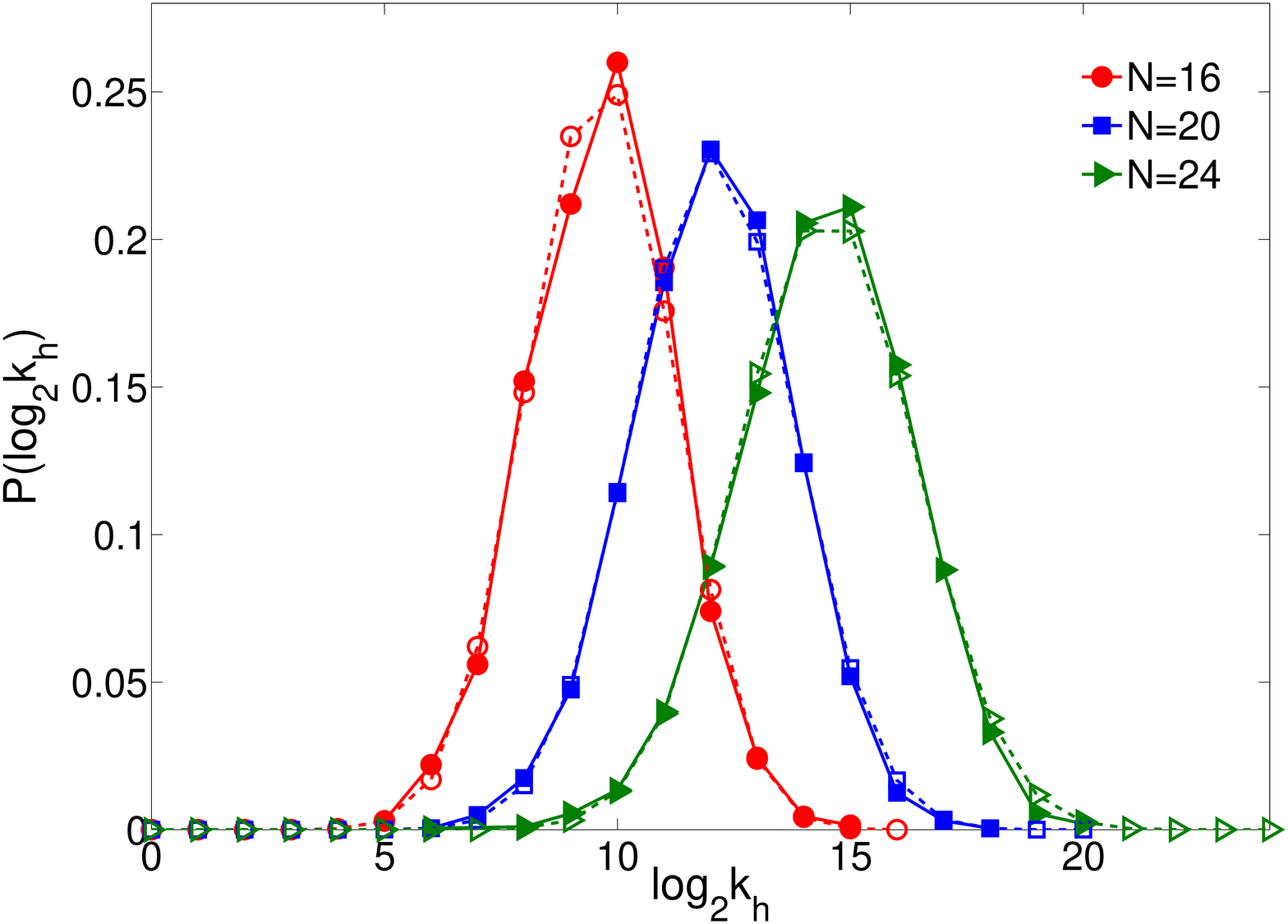}
        \caption{\label{PDFlogkmaxK1}(Color online). The PDF $P(\log_2 k)$
          of the log of the hub in-degree, $\log_2 k_h$ for $K=1$ RBNs. The
          three dashed lines with empty symbols were obtained using
          Eq.~(\ref{dist-marked}) with $q=1/2$.}
\end{center}
\end{figure}

We can use the distribution of $m$ to generate the distribution of ${\rm
  log}_2 k_h = N-m$.  In Figs. ~\ref{PDFlogkmaxK1crit} and ~\ref{PDFlogkmaxK1}
we present the distribution of logarithm of hub in-degrees ${\rm log}_2 k_h$
for the $K=1$ critical and $K=1$ non-critical RBNs respectively. The
hub-degrees were obtained for an ensemble of $2,000$ randomly chosen RBNs with
$10 \leq N \leq 24$. We also show our analytical results obtained from
Eq.~(\ref{dist-marked}) with $q=1$ for the $K=1$ critical RBNs and with
$q=1/2$ for the $K=1$ RBNs.

We can use Eq.~(\ref{dist-marked}) to obtain the mean number of 
effective leaves,
\begin{eqnarray*}
\langle m \rangle_{\tau+1}&\equiv& \sum_{m=0}^{\tau+1} m M_{\tau+1,m}\\
&=& \sum_{m=0}^{\tau+1} m \left[q\left(1-\frac{m-1}{N}\right)M_{\tau,m-1}\right]\\
&+&\sum_{m=0}^{\tau+1} \left[\left(1-q(1-\frac{m}{N})\right)M_{\tau,m}\right],
\end{eqnarray*}
which reduces to,
\begin{eqnarray*}
\langle m \rangle_{\tau+1} = q + \left(1-\frac{q}{N}\right)\langle m \rangle_{\tau}.
\end{eqnarray*}
This recursion gives the solution for the mean number of 
effective leaves $N-\langle m \rangle_N$, which is also the mean of ${\rm
  log}_2 k_h$,
\begin{eqnarray}
\label{K1meanleaves}
\langle {\rm log}_2 k_h \rangle &=& N\left(1-\frac{q}{N}\right)^N\to N e^{-q}~\ {\rm as}~\ N\to\infty.
\end{eqnarray}


\subsubsection{Networks with $K\geq 2$}
\label{K2}
For $K>1$ RBNs the boolean functions are not the simple copy/invert and
constant functions as in the case of $K=1$. For example, consider the truth
table in Table ~\ref{funcegg}. $\sigma_1$ and $\sigma_2$ are inputs to
$\sigma_3$ in a $K=2$ RBN. $\sigma_2$ effects the value of $\sigma_3$ only
when $\sigma_1 = 1$.  Conversely, $\sigma_1$ effects the value of $\sigma_3$
only when $\sigma_2 = 1$. Such interactions between the inputs for $K>1$ RBNs
make it harder to extend the analytical aruguements that we applied to the
$K=1$ case.

\begin{table}[!h]
\begin{center}
\begin{tabular}{|l||l|}
\hline
$\sigma_1$   $\sigma_2$              & $\sigma_3$\\
\hline
0  ~\  0 & 0\\
\hline
0   ~\ 1 & 0\\
\hline
1   ~\ 0 & 0\\
\hline
1   ~\ 1 & 1\\
\hline
\end{tabular}
\caption{\label{funcegg} A function for $K=2$ RBNs}
\end{center}
\end{table}

\begin{figure}[htbp]
\begin{center}
\includegraphics*[width=8cm]{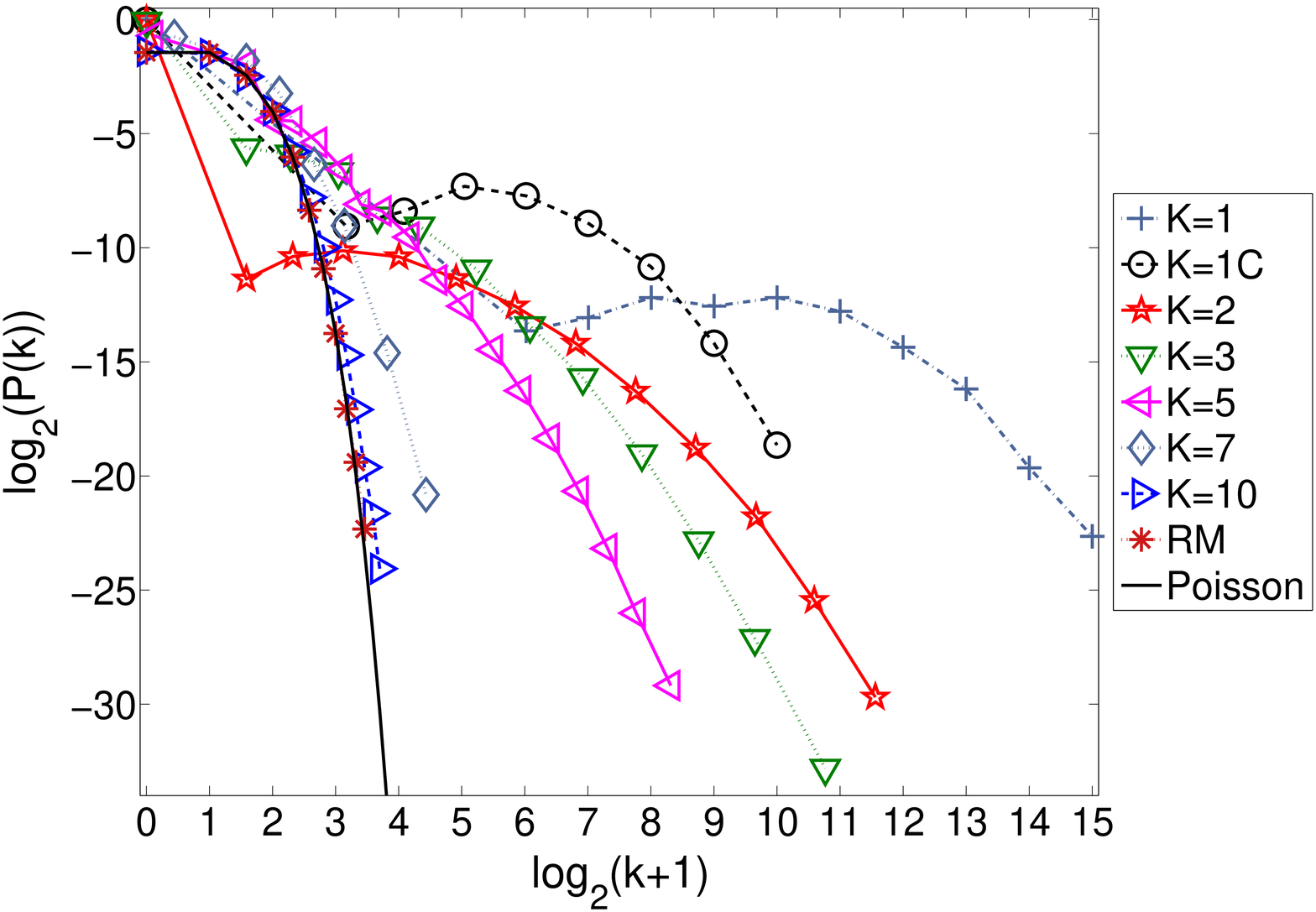}
\caption{\label{PDFInDegN18}(Color online). The in-degree distribution
  function, $P(k)$, for various connectivities, $K$, of the RBN, on a log-log
  scale. The distribution is computed for RBNs of size $N=18$ by averaging the
  PDFs over $200$ different realizations for each $K$. We used $k+1$ instead
  of $k$ on the $x$-axis, so that we could also show values for $k=0$. The
  black solid line shows a Poissonian with $\langle k \rangle = 1$ for
  comparison. $K=1C$ refers to $K=1$ critical RBNs.}
\end{center}
\end{figure}

Fig.~\ref{PDFInDegN18} presents $P(k)$ for SSNs with $N=18$ (${\cal N} =
2^{18}$).  It is broad for $K = 1$, becomes narrower with increasing $K$, and
converges for $K\to N$ to the in-degree distribution of the random map.  The
random map in-degree distribution itself converges to a Poissonian with
$\langle k \rangle = 1$ when $N \to \infty$.  Note that $k$ for $K=1$ is
either zero for the GoE states, or takes the same power of two value for the
hubs.

\begin{figure}[htbp]
\begin{center}
\includegraphics*[width=8cm]{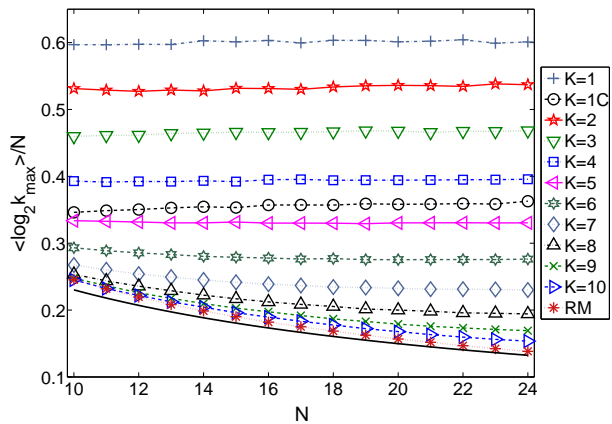}
\begin{center}
(a)
\end{center}
\includegraphics*[width=8cm]{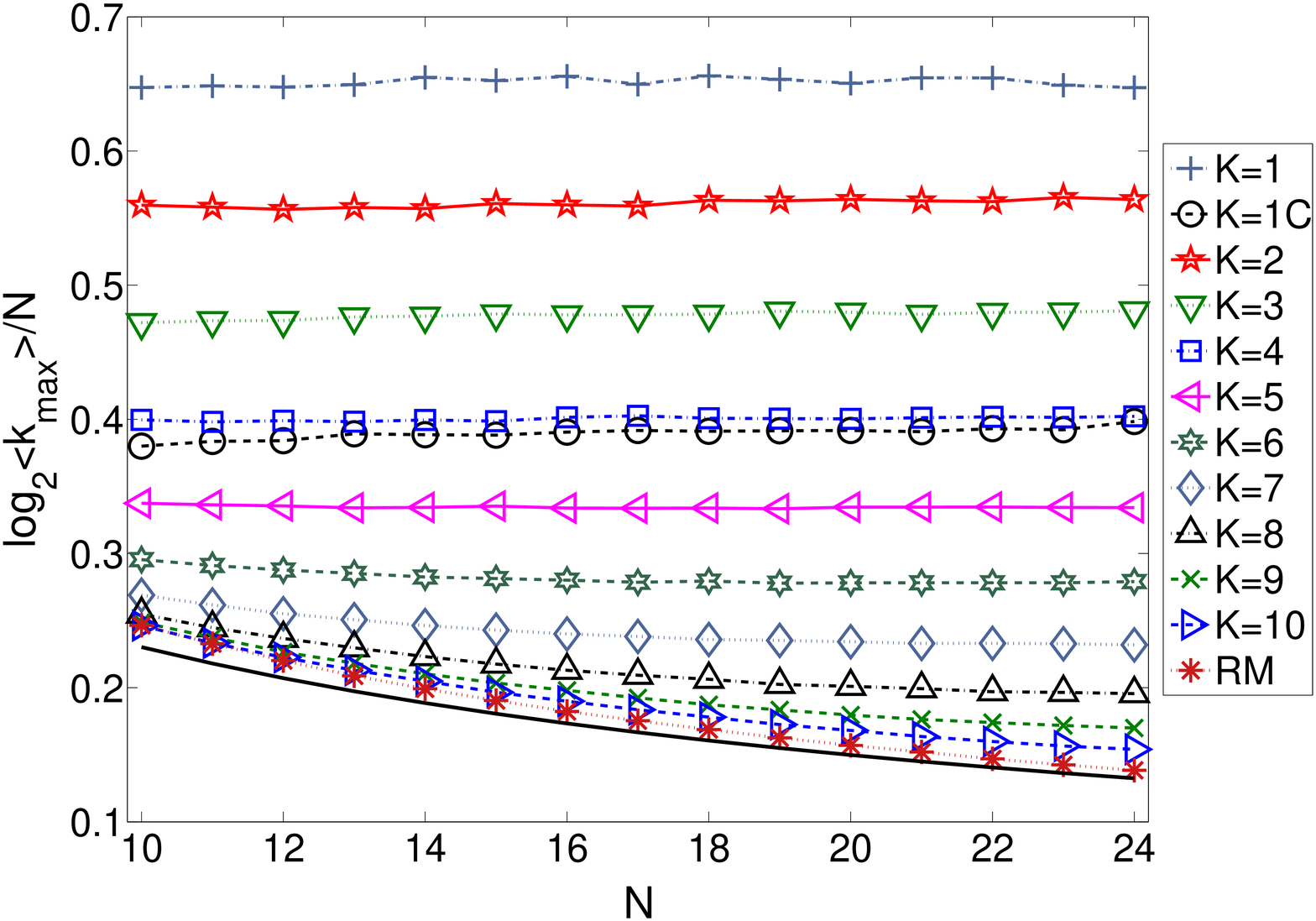}
\begin{center}
(b)
\end{center}
\caption{\label{checklinearmeanlogkmax}(Color online). (a) The ratio
  $\frac{\langle \log_2 k_{max} \rangle}{N}$ is plotted against $N$. (b) The
  ratio $\frac{\log_2 \langle k_{max} \rangle}{N}$ is plotted against $N$.
  Both figures show systematic deviations from scaling for $K>6$.  The solid
  lines in (a) and (b) are $ \frac{\log N}{N} $ and were plotted for
  comparison. The statistics, in this figure and
  Fig.~\ref{meanlogkmaxDivlogmeankmax} were obtained by sampling $2000$
  different realizations for each $N$ and $K$.}
\end{center}
\end{figure}

The in-degree distribution $P(k)$ does not exhibit easily described behavior
such as scaling.  However, the results of Ref.~\cite{Shreim2007CA} suggest
that a more useful quantity is the largest in-degree $k_{max}$.  Scaling
behavior in $k_{max}$ was shown to be a necessary but not sufficient signature
of complex dynamics.  Fig.~\ref{checklinearmeanlogkmax}a shows that
\begin{equation}
  \langle \log_2 k_{max} \rangle \sim \nu_K N {~\rm for } ~\ K \leq 6,
         \label{eq2}
\end{equation}
where the angular brackets indicate an average over different realizations of
the RBN.  The exponents $\nu_K$ appear to obey the relation:
\begin{equation}
\nu_K = -0.07 K + 0.68 {~\rm for } ~\ 1 \leq K \leq 6,
\end{equation}
where the error is $\pm 1$ in the last digit. For the system sizes studied, it is not
possible to determine the asymptotic limit for larger $K$. However, the
behavior for large $K$ approaches the random map result.

For the random map, $P(k)$ tends to a Poisson distribution when $N \to
\infty$. If $\cal N$ different values of $k$ are chosen from a distribution
$P(k)$, the expected maximum $k_{max}$ can be related to $\cal N$ by
\begin{equation}
\label{kmaxrmeq}
\sum_{k=k_{max}}^{\infty} P(k) = \frac{1}{\cal N}.
\end{equation}
For a Poissonian with mean one,
\begin{eqnarray*}
\sum_{k=k_{max}}^{\infty} \frac{e^{-1}}{k!} = \frac{1}{\cal N}.
\end{eqnarray*}
This gives the following bounds on $k_{max}$:
\begin{eqnarray*}
  \frac{e^{-1}}{k_{max}!} \leq \frac{1}{\cal N} \leq \frac{e^{-1}}{k_{max}!} \sum_{m=0}^{\infty} \frac{1}{(k_{max})^{m}}. 
\end{eqnarray*}
The sum in the last term in the equation evaluates to $1 - 1/k_{max}$ which
goes to one for large $k_{max}$, allowing us to write
\begin{eqnarray*}
\frac{e^{-1}}{k_{max}!} \approx \frac{1}{\cal N}. 
\end{eqnarray*}
Taking logarithms and using Stirling's approximation, the above equations
yields for large ${\cal N}$
\begin{equation}
  k_{max}\ln k_{max} \approx  \ln{\cal N} = N\ln 2.
\end{equation}
Finally we take logarithms again to get,
\begin{equation}
  \log_2 k_{max} \approx  \log_2 N.     \label{kmaxN}
\end{equation}

As shown in from Fig.~\ref{checklinearmeanlogkmax}a, $\langle \log_2 k_{max}
\rangle$ is very well described by Eq.(\ref{kmaxN}) for random maps and RBNs
with large $K$ in the limit $N\to\infty$.

\begin{figure}[htbp]
\begin{center}
\includegraphics*[width=8cm]{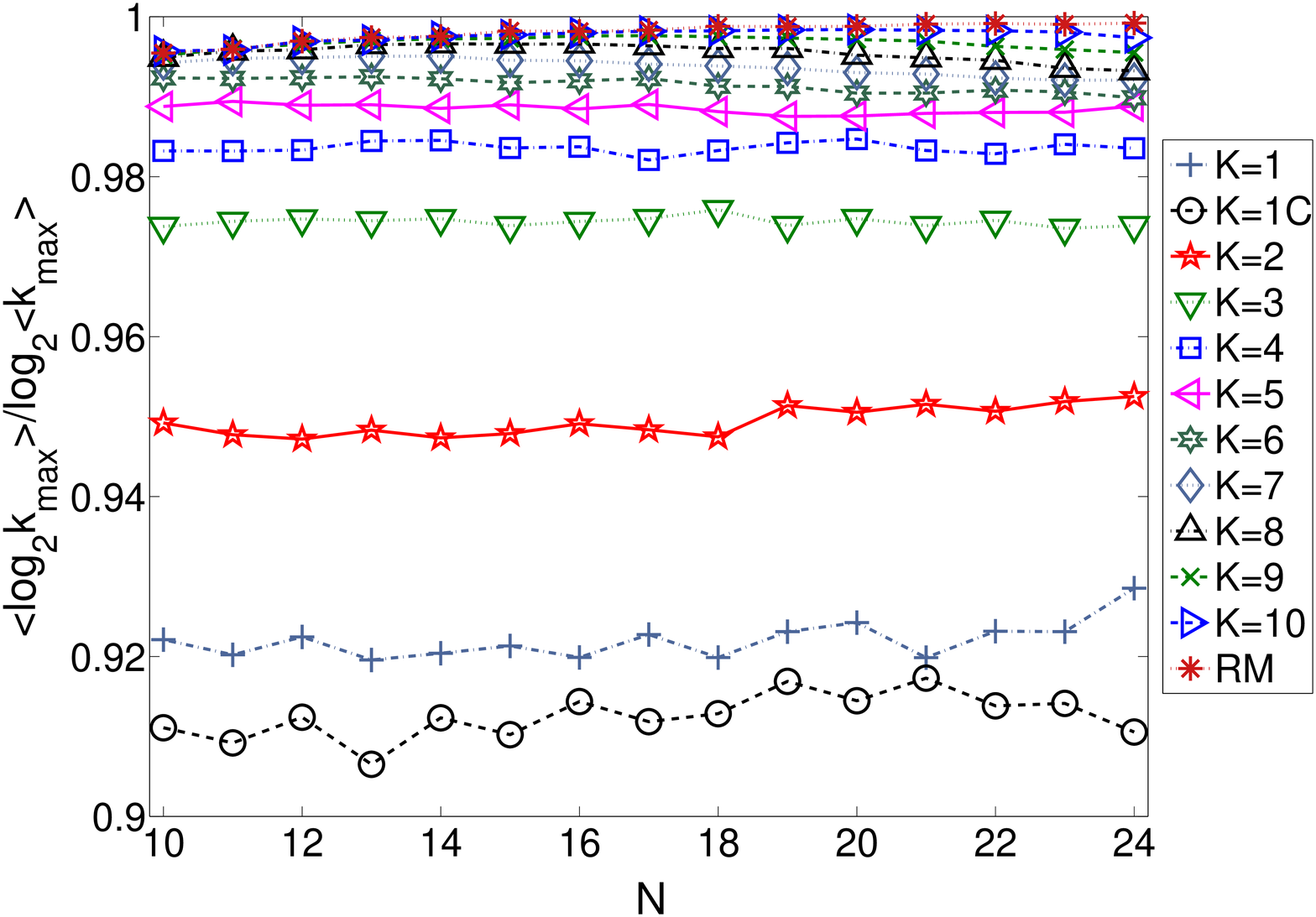}
\caption{\label{meanlogkmaxDivlogmeankmax} (Color online). The ratio between
  the mean of the log and the log of the mean of the largest in-degree,
  $y\equiv \frac{\langle \log_{2} k_{max} \rangle}{\log_{2} \langle k_{max}
    \rangle}$, is plotted against the size of the RBN, $N$. If the two
  quantities were scaling with the same exponent, all curves should tend to
  $y=1$ for large $N$. The plot shows that $\langle \log_{2} k_{max} \rangle$
  and $\log_{2} \langle k_{max} \rangle$ are scaling with different
  exponents for $K \leq 6$.}
\end{center}
\end{figure}

In order to see whether $k_{max}/N$ is self averaging, {\it i.e.} whether the
fluctuations of $k_{max}/N$ become negligible for large $N$, we also
plot in Fig.~\ref{checklinearmeanlogkmax} the ratio $N^{-1}\log_2 \langle
k_{max} \rangle$. A first look at Fig.~\ref{checklinearmeanlogkmax} might
suggest that both ways of averaging lead indeed to the same results, and
self-averaging is satisfied. That this is not the case is demonstrated in
Fig.~\ref{meanlogkmaxDivlogmeankmax}, where we show the ratio $\langle \log_2
k_{max} \rangle/\log_2\langle k_{max} \rangle$ versus $N$.  We see from this
figure that $\langle k_{max} \rangle$ scales with ${\cal N}$ as $\langle
k_{max} \rangle \sim {\cal N}^{\tau_K}$ for $K \leq6$, {\it i.e.}  $\log_2
\langle k_{max}\rangle \sim \tau_K N$, similarly to Eq.(\ref{eq2}). However,
$\tau_K \neq \nu_K$. For large $K$ it seems that that the fluctuations are
less important, $\langle \log_2 k_{max} \rangle/\log_2\langle k_{max} \rangle
\to 1$ for $K\to\infty$.

\begin{figure}[htbp]
\begin{center}
\includegraphics*[width=8cm]{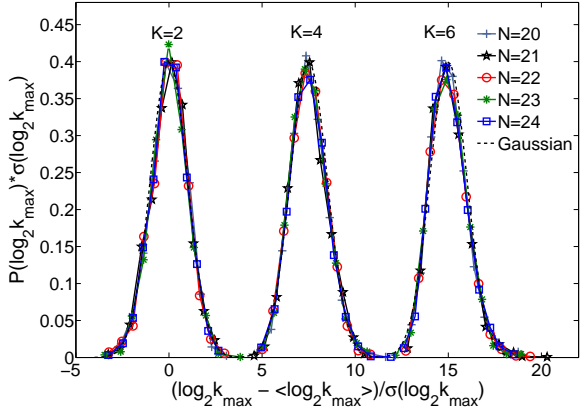}
\begin{center}
(a)
\end{center}
\includegraphics*[width=8cm]{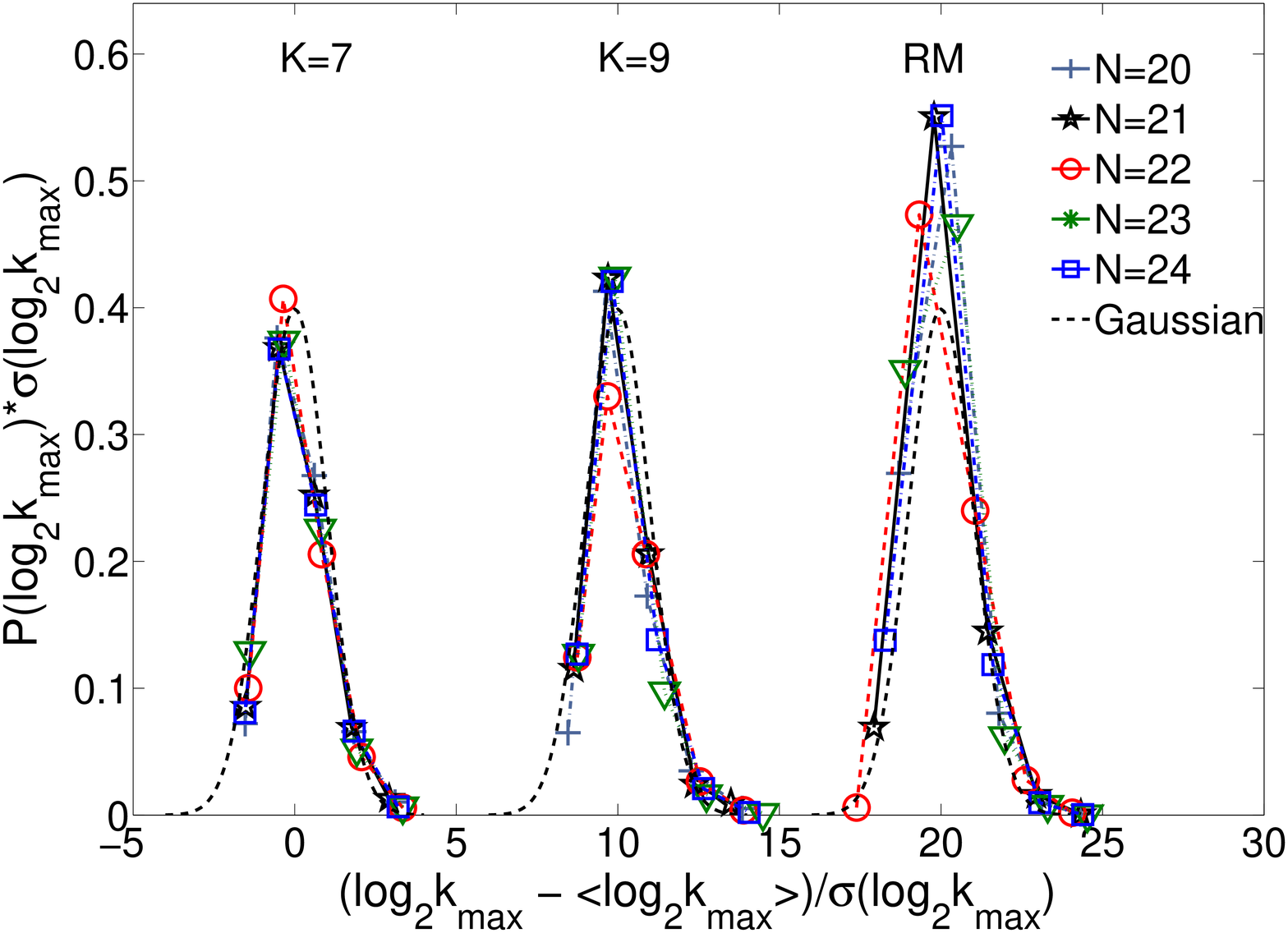}
\begin{center}
(b)
\end{center}
\caption{\label{RescaledPDFkmaxK246}(Color online). The rescaled PDFs of the
  log of the largest in-degree $P(\log_2 k_{max})$ for (a)$K=$ 2, 4, and 6,
  and (b) $K=$ 7, 9, and the random map. The dashed lines are Gaussian
  distributions with mean zero and variance one. The plots indicate that
  $k_{max}$ is distributed according to a log normal distribution for $K=$ 2,
  4 and 6. Similar results hold for $K =$ 3 and 5.  Deviations from the
  Gaussian distribution are seen for $K>6$.  In (a) the distributions for
  $K=4$ and $K=6$ were offset by $7.5$ and $15$ units on the x-axis for
  clarity of presentation. In (b) the distributions for $K=9$ and the random
  map were offset by $10$ and $20$ units.}
\end{center}
\end{figure}

The sample-to-sample fluctuations in $k_{max}$ are captured by its probability
distribution, shown in Fig.~\ref{RescaledPDFkmaxK246}.  We plot the
probability distribution of $\log k_{max}$ multiplied by its standard
deviation, $\sigma(\log_2 k_{max})$. The mean and standard deviation were
computed independently for each $N$ and $K$.  The plots are well-fitted by a
Gaussian for $2 \leq K \leq 6$, which suggests that $P(k_{max})$ is a
log-normal, for sufficiently large $N$. For $K>6$, the plots indicate
deviations from log-normal behavior.

\subsection{Path Diversity} 
\label{diversity}

Scaling of $k_{max}$ with system size for $K\leq6$, indicates asymptotic local
heterogeneity of the SSNs. As shown in Ref.~\cite{Shreim2007CA} local
heterogeneity is often insufficient to distinguish simple from complex
dynamics.  A complimentary global topological measure, the path diversity,
detects global heterogeneity in SSNs. In Ref.~\cite{Shreim2007CA} it was found
that complex cellular automata show scaling behavior in both local and global
measures. Here we investigate the path diversity of SSNs of RBNs.

\begin{figure}[htbp]
\begin{center}
\includegraphics*[width=8cm]{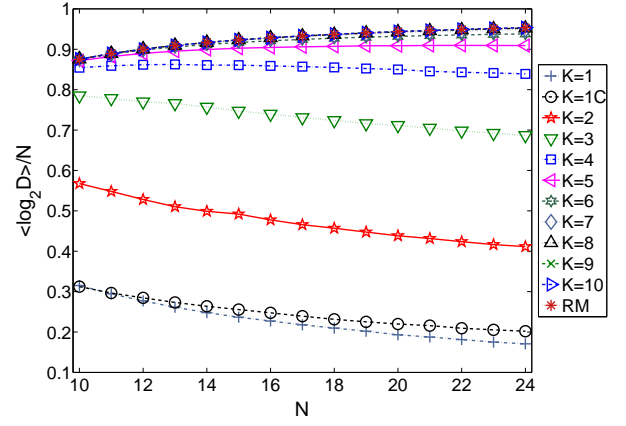}
\caption{\label{meanlogD} (Color online). Log path diversity as a function of
  the system size. $\langle \log_2{\cal D} \rangle/N$, is plotted as a
  function of $N$, for various values of $K$ and the random map. For $K \geq 5$,
  $\langle \log_2{\cal D} \rangle$ asymptotes to $\zeta_K N$, while the plots
  have the opposite curvatures for $K\leq4$.}
\end{center}
\end{figure}

\begin{figure}[htbp]
\begin{center}
\includegraphics*[width=8cm]{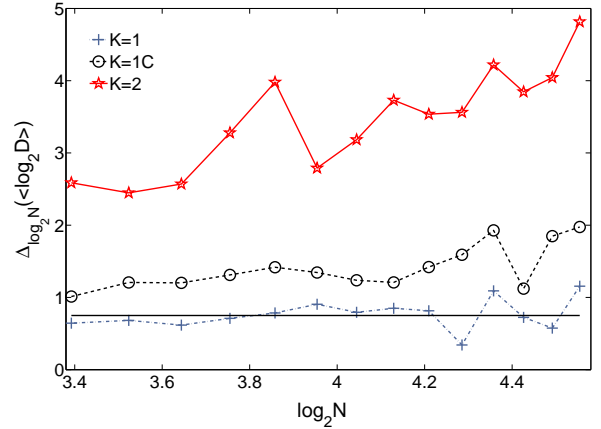}
\caption{\label{DerivmeanlogDVslogN}(Color online). The discrete derivative of
  $\langle \log_2{\cal D} \rangle$ with respect to $\log_2N$,
  $\Delta_{\log_2N} \langle \log_2{\cal D} \rangle$, is plotted against
  $\log_2N$ for $K = 1,2$ and $K = 1$ critical. $\Delta_{\log_2N} \langle
  \log_2{\cal D} \rangle$ is defined as $\frac{ \langle \log_2{\cal D}_i
    \rangle - \langle \log_2{\cal D}_{i-1} \rangle}{\log_2N_i -
    log_2N_{i-1}}$.  A horizontal line of this plot indicates a relationship
  of the form $\langle \log_2{\cal D} \rangle \sim \log_2 N$. For the system
  sizes studied, the data suggests a logarithmic growth of $\langle
  \log_2{\cal D} \rangle$ with respect to $N$. For $K=2$, $\Delta_{\log_2N}
  \langle \log_2{\cal D} \rangle$ show growth trends which are indicative of a
  faster than logarithmic growth of $\langle \log_2{\cal D} \rangle$ with
  $N$. The data is inconclusive for $K=1$ critical.}
\end{center}
\end{figure}

Fig.~\ref{meanlogD} suggests that $\langle \log_2{\cal D} \rangle \sim \zeta_K
N$ for $K \geq 5$ and for the random map. However, the plots are not
conclusive as to whether or not $\langle \log_2{\cal D} \rangle$ becomes
linear in $N$ for $K<5$, for large system sizes, or with large finite size
corrections. Fig.~\ref{DerivmeanlogDVslogN} shows $\Delta_{\log_2N} \langle
\log_2{\cal D} \rangle$, the discrete derivative of $\langle \log_2{\cal D}
\rangle$ with respect to $\log_2N$. For $K=1$, the figure indicates that
$\langle \log_2{\cal D} \rangle \sim \log_2 N$. A faster than logarithmic
growth of $\langle \log_2{\cal D} \rangle$ is seen for $K=2$.

An example of a simple tree that exhibits logarithmic behavior of path
diversity is the Cayley tree. On such an SSN each transient state has exactly
$z$ pre-images, except for the root which has $z+1$ pre-images. Given the
symmetry of the Cayley tree, all the nodes at the same distance from the root
will have the same path diversity. Furthermore, the path diversity is
incremented by one with each step towards the root. This makes the path
diversity equal to the depth of the tree which can be expressed in terms of
the number of nodes ${\cal N}$:
\begin{equation}
\label{diversityTree}
{\cal D} = \log_z\{(z -1){\cal N} + 1\}.
\end{equation}

For a maximally heterogeneous SSN each transient node will contribute to the
path diversity. The path diversity of an SSN is the diversity of the root.
Diversity of a node is the sum of all distinct diversities of its pre-images
plus one if it has more than one pre-image. Thus the diversity of a node is
bounded above by,
\begin{eqnarray*}
{\cal D}_i \leq 1+ \sum_{j \prec i} {\cal D}_j
\end{eqnarray*}
where $j \prec i$ indicates that $j$ is a pre-image of $i$. In the general
case, only distinct values of the diversities will contribute to the sum on
the right hand side. We can iteratively use this upper bound to calculate the
path diversity of an SSN. Denoting the root of the SSN by $\bullet$,
\begin{eqnarray*}
{\cal D}&\leq& 1 + \sum_{i_1 \prec \bullet}{\cal D}_{i_1}\\
&=&1 + \sum_{i_1 \prec \bullet} 1 + \sum_{i_2 \prec i_1}\sum_{i_1 \prec \bullet}{\cal D}_{i_2}\\
&=&1 + \sum_{i_1 \prec \bullet} 1 + \sum_{i_2 \prec i_1}\sum_{i_1 \prec \bullet} 1 +  \sum_{i_3 \prec i_2}\sum_{i_2 \prec i_1}\sum_{i_1 \prec \bullet}{\cal D}_{i_3}
\end{eqnarray*}
and so on. The subscript $h$ in $i_h$ represents the distance of the node
$i_h$ from the root. If $i_h$ is a GoE state, its diversity is defined to be
one. The first term on the right hand side of the last line counts the root,
the first sum counts the root's pre-images, and the two nested sums count the
pre-images of the root's pre-images. Iteratively continuing this calculation
we can see that the right hand side evaluates to ${\cal N}$, the size of the
SSN. Thus we have as the upper bound to the path-diversity,
\begin{equation}
{\cal D}\leq {\cal N}=2^N,
\end{equation}
or that,
\begin{equation}
{\rm log_2} {\cal D}\leq N.
\end{equation}

This bound is nearly reached for binary tress where single leaves branch off
at every branching point. 

The logarithmic growth of $\langle \log_2{\cal D} \rangle$ for $K=1$ networks
suggests a different class of dynamical complexity than for $K\geq 2$
networks.  Meanwhile, random map-like-scaling of $\langle \log_2 k_{max}
\rangle$ for RBNs with sufficiently large $K$ suggests relatively simple
dynamics.  However, on the basis of those two measures, one cannot cleanly
distinguish the behavior of $K=2$ critical RBN from that of $K>2$ RBNs.  We
now show that sample-to-sample fluctuations in $\log_2{\cal D}$ enable such a
distinction.

\begin{figure}[htbp]
\begin{center}
\includegraphics*[width=8cm]{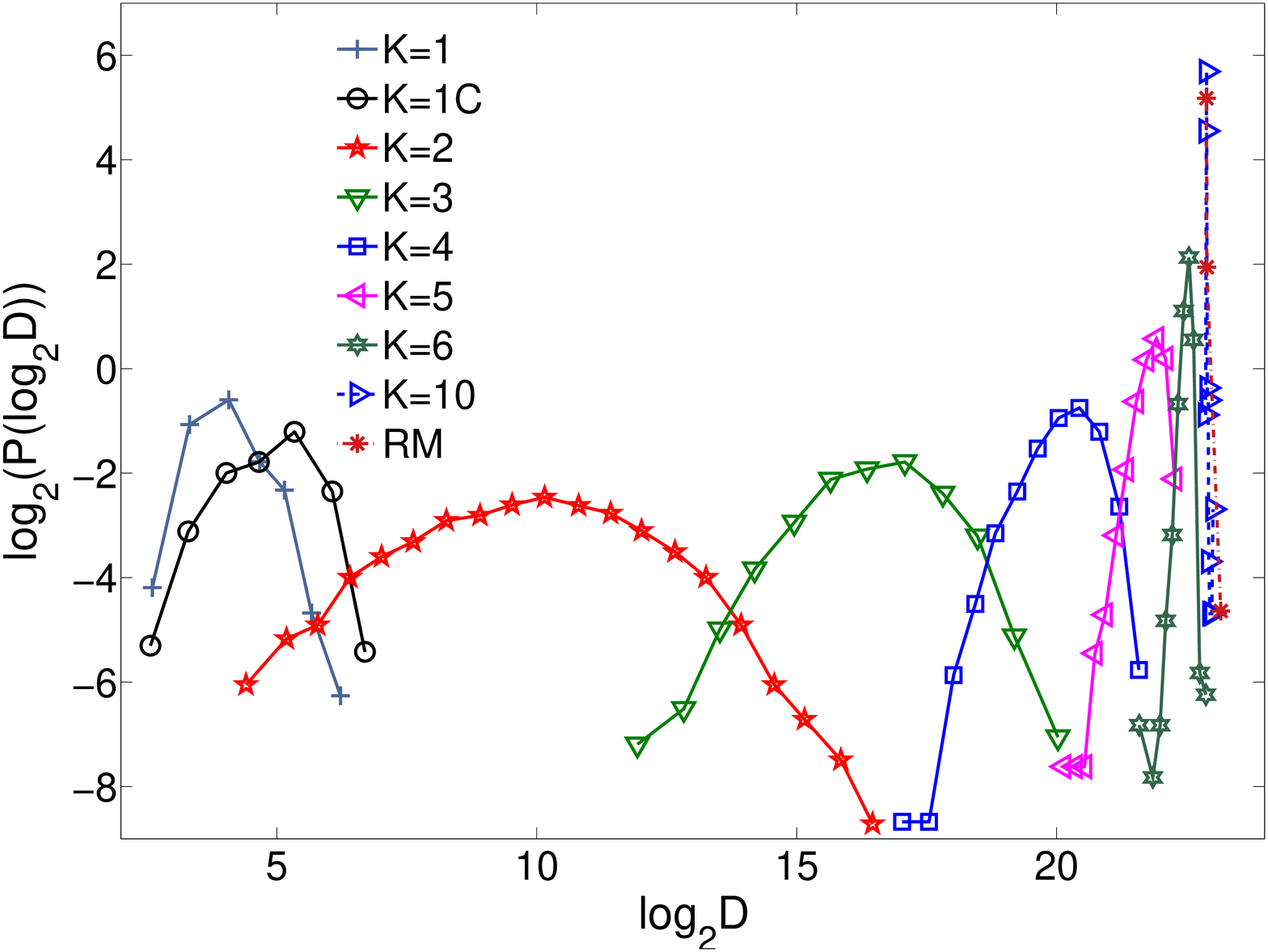}
\caption{\label{PDFlogD} (Color online). The log of the PDF of the log of the
  path diversity, $P(\log_2{\cal D})$, for various values of $K$ and $N=24$.
  $P(\log_2{\cal D})$ is the broadest for $K=2$. $P(\log_2{\cal D})$ becomes
  narrow for large values of $K$.}
\end{center}
\end{figure}

\begin{figure}[htbp]
\begin{center}
\includegraphics*[width=8cm]{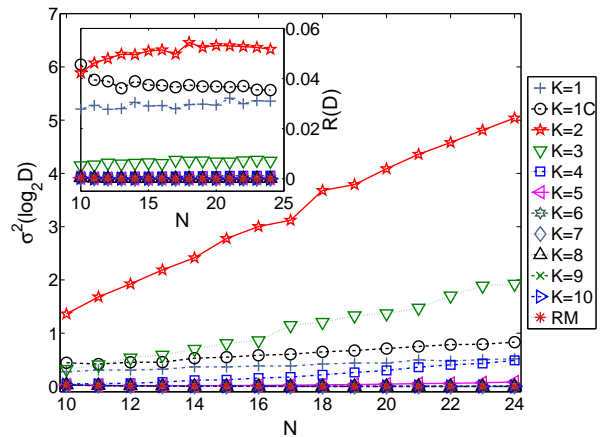}
\caption{\label{varlogD} (Color online). The variance of the log of the path
  diversity $\sigma^2(\log_2{\cal D})$ is plotted as function of the size of
  the RBN, $N$, for various values of $K$. $\sigma^2(\log_2{\cal D})$ shows
  non-monotonic behavior as a function of $K$, it grows the fastest for $K=2$.
  On the other hand, $\sigma^2(\log_2{\cal D})$ appears to tend to a constant
  or decreases with increasing $N$, for large values of $K$. For the random
  map, $\sigma^2(\log_2{\cal D})$ is a decreasing function of $N$. The inset
  shows a plot of ${\cal R}({\cal D}) = \frac{\sigma^2(\log_2{\cal
      D})}{\langle \log_2{\cal D} \rangle ^2}$ as function of $N$. The data
  for $K=2$ suggests that $\log_2{\cal D}$ might be non-self-averaging in the
  thermodynamic limit.}
\end{center}
\end{figure}

We report the distribution of $\log_2 {\cal D}$, $P(\log_2{\cal D})$, in
Fig.~\ref{PDFlogD}.  For a fixed system size, $P(\log_2{\cal D})$ is broadest
for $K=2$ RBNs.  To quantify the width of $P(\log_2{\cal D})$, we study its
variance.  Fig.~\ref{varlogD} shows that the variance of $\log_2{\cal D}$,
$\sigma^2(\log_2{\cal D})$, grows fastest with $N$ for $K=2$.  Unlike $\langle
\log_2 k_{max} \rangle$ and $\langle \log_2{\cal D} \rangle$,
$\sigma^2(\log_2{\cal D})$ shows non-monotonic behavior as a function of $K$.
We interpret this as an indication that the criticality of $K=2$ RBNs can be
associated with large sample-to-sample fluctuations of its SSNs.

Generally, large sample-to-sample fluctuations in disordered media may lead to
the absence of self-averaging.  A canonical measure of self-averaging is the
ratio:
\begin{equation} 
   {\cal R}({\cal D}) = \frac{\langle {\cal D}^2 \rangle - \langle {\cal D}
     \rangle^2}{\langle {\cal D}  \rangle^2}. 
 \end{equation}
 A system is said to be self-averaging with respect to ${\cal D}$ if ${\cal
   R}({\cal D})$ goes to zero for large $N$~\cite{chamati2002fss,
   wiseman1998fss}; otherwise, it is said to lack self-averaging.  For the
 system sizes we are able to study, the evidence for or against self-averaging
 is not completely conclusive. Nevertheless, as indicated in the inset of
 Fig~\ref{varlogD}, $K=2$ RBNs show the largest values for ${\cal R}$ and are
 therefore the most likely to exhibit non-self-averaging behavior in the
 thermodynamic limit of large system size.

\section{Discussion and Conclusion}
\label{discussion}
We have studied the topology of state space networks (SSNs) for ensembles of
random Boolean networks.  Each dynamical state of the Boolean network
corresponds to a node in the SSN and is linked to its successor state.  We
characterize the heterogeneity of these SSNs at the local, node, scale by the
distribution of the in-degrees and the scaling of the largest in-degree.
Global heterogeneity over all paths in the SSN is characterized by the path
diversity. For elementary 1-d cellular automata, it was demonstrated in
Ref.~\cite{Shreim2007CA} that simultaneous scaling behavior in both $k_{max}$
and ${\cal D}$ indicates ``complex'' spatio-temporal dynamics (Wolfram class
IV and partly class III).  On the other hand, it was found in
\cite{Shreim2007CA} that one or both of these does not scale for CA in class I
or II -- which all have simple dynamics.
 
As is well-known, RBNs exhibit a phase transition between chaotic ($K> 2$) and
frozen ($K<2$) behavior.  $K=2$ RBNs are critical and therefore more complex
than RBNs with $K\neq 2$ which mainly show frozen ($K<2$) or chaotic ($K>2$)
behavior~\cite{OriginOrder}.  In fact, RBNs in the frozen phase (e.g. $K=1$)
resemble class II CA as they almost always rapidly go to one of many
attractors with a short period.  In addition, RBNs in the chaotic phase
resemble some class III CA in that they have long transients and attractors
with large periods.  We have investigated whether or not the different phases
of RBNs can be distinguished on the basis of a topological analysis of the
corresponding ensembles of SSNs, and the extent to which these phases overlap
in the behavior of their SSNs.

Many analytical results are known for $K=1$
RBNs~\cite{flyvbjerg1988esk,Drossel2005PRL} (and also for the random map,
\cite{DerridaFlyvbjerg}). In the present paper we derived analytical results
for the structure of the state space network for $K=1$ RBNs, and verified them
using numerical methods. We related the hub in-degree $k_h$ to $l$, the number
of effective leaves in the RBN. Simple arguments show that all the hubs have
the same in-degree $k_h = 2^l$. We also presented a stochastic process in
Eq.~(\ref{dist-marked}) that generates the distribution of $l$ for any $K=1$
RBN. For $K=1$ critical RBNs we could obtain an analytical expression for the
distribution of the hub in-degree $k_h$, and compared numerical solutions to
Eq.~(\ref{dist-marked}) with the simulations for the general $K=1$ RBNs.

Using Eq.~(\ref{dist-marked}) we derived that for $K=1$, ${\rm log}_2 k_h$
scales with the size of the RBNs as $\langle{\rm log}_2 k_h\rangle = e^{-q}N$,
where $q$ is the fraction of non-constant functions.  Numerically we find that
for $2 \leq K \leq 6$, $\log_2 k_{max}$ also exhibits scaling as a function of
the size of the RBN, $\langle \log_2 k_{max} \rangle \sim \nu_K N$.  The
scaling exponent $\nu_K$ is largest for $K=1$ and monotonically decreases for
larger values of $K$.  For $K=1$ the hub in-degree $k_h$ depends exponentially
on the number of effective leaves, $l$, which scales linearly with $N$. The
state of effective leaves is forgotten after the first time step during
evolution of the RBN. For $K\geq 2$, there are more than one inputs to each
element.  Whether the state of an element is forgotten or not depends on the
state of other elements in the RBN. Observed linear scaling of ${\rm log}_2
k_{max}$ with $N$ for $2\leq N \leq 6$ indicates an explanation similar to the
one we presented for $K=1$.  However, more analytical work is required to
fully understand the relationship between $k_{max}$ and the structure of the
RBN for $K>1$.  For larger $K$, we notice that the $K,N \to \infty$ limit of
the RBNs is the random map~\cite{DerridaFlyvbjerg}. We have shown analytically
that for the random map, $\log_2 k_{max} \sim \log_2 N$.

On the other hand, our numerical results indicate that $\langle \log_2{\cal D}
\rangle$ scales with the size of the SSN only for sufficiently large values of
$K$ ($K \geq 5$), $\langle \log_2{\cal D} \rangle \sim \zeta_K N$. $\zeta_K$
is largest for the random map and monotonically decreases with decreasing $K$
(which is the opposite of $\nu_K$).  The logarithmic scaling with $N$ of
$\langle \log_2{\cal D}\rangle$ for $K=1$ translates to a logarithmic scaling
of ${\cal D}$ with the size of the SSN, ${\cal N}$. This compares with the
path diversity of a Cayley tree in Eq.~(\ref{diversityTree}).  On an SSN with
Cayley tree structure, all nodes, except GoE nodes, have the same in-degree.
In fact we have shown the same for $K=1$ SSNs.  For $ 2 \leq K \leq 5$, it is
not clear from the data whether $\langle \log_2{\cal D} \rangle$ scales
linearly with $N$.  It is clear, however, that $\langle \log_2{\cal D}
\rangle$ grows faster-than-linear with $\log_2{N}$.

Together, the absence of scaling for $k_{max}$ and ${\cal D}$ correctly rule
out the random map (and most likely RBNs with large $K$) and $K=1$ RBNs, from
reporting high dynamical complexity.  However, neither of these measures
addresses sample-to-sample fluctuations, which are generally important in the
characterization of disordered systems.  We find that the probability
distribution function of $k_{max}$ converges to a log-normal distribution for
$2\leq K \leq 6$.  While this is also seen for $\log_2{\cal D}$, unlike the
monotonic dependence of the variance of $\log_2 k_{max}$, the variance of
$\log_2{\cal D}$ does not vary monotonically with $K$. In fact, the variance
grows the fastest with the size of the SSN for critical $K=2$ RBNs.  Numerical
results also suggest that $\log_2{\cal D}$ may possibly exhibit
non-self-averaging behavior for $K=2$.

Our results support the conclusion that heterogeneity in SSNs can be
associated with complex dynamics in disordered systems. $K=2$ RBNs are
distinguished from other RBNs by simultaneously exhibiting three kinds of
network heterogeneity.  The first is a local heterogeneity on the node level
indicated by the scaling of $\log_2 k_{max}$ with the size of the SSN.  The
second is a global heterogeneity on the trajectories level indicated by a
faster-than-linear growth of $\log_2{\cal D}$ with $\log_2N$. Finally, there
is a heterogeneity on the level of samples of RBNs, indicated by the fast
growth of the $\sigma^2(\log_2{\cal D})$ for RBNs with $K=2$.

\begin{acknowledgments}
  We would like to thank one of the referees for suggesting a way to prove
  that the in-degrees of a $K=1$ RBN only comes as a power of two. VS would
  like to thank Fanny Dufour for insightful discussions.
\end{acknowledgments}

\bibliography{references}

\end{document}